\documentstyle[aps]{revtex}
\twocolumn

\begin{document}

\title{Invariant information and quantum state estimation}

\author{J. \v{R}eh\'{a}\v{c}ek and Z. Hradil}
\address{Department of Optics, Palacky University, 17. listopadu 50,
772 00 Olomouc, Czech Republic}
\date{\today}
\maketitle


\begin{abstract}
The invariant information introduced by Brukner and Zeilinger,
{\em Phys. Rev. Lett. 83, 3354 (1999),}
is reconsidered from the point of view of quantum state estimation.
We show that it is directly related to the mean error of the standard
reconstruction from the measurement of a complete set of mutually
complementary observables. We give its generalization in terms of
the Fisher information. Provided that the optimum 
reconstruction is adopted, the corresponding 
quantity loses its invariant character. 
\end{abstract}

\vspace{0.5cm}

Invariants are important concepts of physics, because referring to
them, physicists build up theoretical models making it possible to
explain consistently their observations.
Recently a project has been undertaken to
build up the kinematics of the quantum theory
from the information-theoretical principles
\cite{zeilinger,caslav,malus,essence}.
As a part of the project a new measure of the classical
information gained from a measurement on a quantum system
was introduced. This quantity summed over a complete set of
mutually complementary observables (MCO) exhibits invariance with
respect to unitary transformations applied to the
state of the system and/or to the measured set of MCO.
Moreover, when properly normalized, it also quantifies the 
(maximum) information content, and therefore evaluates the 
processing power of physical systems.

A nice feature of the invariant information of Brukner and Zeilinger
is that its definition is operational. It is obtained
by synthesizing the errors of a specially chosen set of
measurements performed on the
system. In this contribution, we analyze the invariant information
from a different perspective. If certain observations are made on
the system the obtained results can be in a natural way put
together to form our estimate of the quantum state of the 
system.
This hints on the existence of a tight link between the
information gained from a particular  set of  measurements and
the error of the reconstruction based on the
obtained results. Being motivated by the estimation theory
we will show how to synthesize information
gained from individual measurements in more general situations,
namely, when (i) a complete set of MCO is not available, and 
when (ii) the  observables
measured are not necessarily mutually complementary
(still non-commutative but not ``maximally'' non-commutative).
Special attention will be paid to the invariance properties of the total
information gained from a measurement.
As a by-product, we will show an alternative interpretation of the
invariant information and will discuss the role that MCO play 
in the state estimation.

Let us consider a measurement of an observable 
$A=\sum_{j=1}^{p} a_j \Pi_j$
with a non-degenerated spectrum acting in the Hilbert space of dimension
$p$. Each outcome $j$ is detected with probability 
$p_j={\rm Tr}\rho \Pi_j$.
Brukner and Zeilinger's lack of information associated with such a measurement
is defined as a sum of variances of individual outcomes per particle
$E_A=\sum_j\sigma_j^2/N$, where $N$ is the number of particles
incident on the measurement apparatus \cite{caslav}. 
The total lack of information
about the system is then obtained as a sum of those measures
over a complete set of MCO,
 \begin{equation} \label{error}
E=\sum_{\alpha j}\sigma^2_{\alpha j}/N=
\sum_{\alpha j}p_{\alpha j}(1-p_{\alpha j}),
\end{equation}
where $\alpha=1\ldots p+1$ and $j=1\ldots p$ \/ label complementary
observables and their eigenvectors, respectively.
The invariant information is nothing else than a properly normalized
complement of the total lack of information $E$ \cite{caslav}.
Assume that a complete set of MCO exists for the given dimension $p$,
which means there are $p(p+1)$ projectors $\Pi_{\alpha j}$
satisfying \cite{ivanovic,schwinger}
\begin{equation} \label{def-compl}
{\rm Tr}\{\Pi_{\alpha j}\Pi_{\beta k}\}=
\delta_{\alpha\beta}\delta_{jk}+1/p(1-\delta_{\alpha\beta}).
\end{equation}
One easily finds that the total error has already the desired invariance
under the choice of the complete set of mutually complementary measurements,
and under unitary transformation of the true state $\bar{\rho}$ of 
the system:
\begin{equation} \label{invar}
E=p-{\rm Tr}\bar\rho^2.
\end{equation}
The main argument for just summing up the errors of complementary
observations is that MCO are independent in the sense that a
measurement of one of them gives no information about the 
rest \cite{wf-mco}.
This is the standard interpretation of MCO observables based on the
following statistical reasoning: Equation (\ref{def-compl}) could be
interpreted as the conditional probability $p(\alpha,j|\beta,k)$
corresponding to the detection of the quantum variable $\{ \alpha,j\}$
provided that $\{ \beta,k\}$ was true. This is symmetrical with
respect to the interchange of observation and conditioning.
Consequently, if a particular  observable $\alpha$ is measured,  
the Bayesian posterior distributions associated with the observables
$\beta\neq \alpha$ complementary to it are flat. Such
observation seems to enhance our knowledge just about a particular
detected output without adding anything new to the unobserved
variables. For this reason, MCO are sometimes
called ``unbiased observables'' \cite{positivity}. 
As will be shown in the following 
such a classical interpretation of MCO variables should  be modified.

There is a close link between the total error $E$ and the error of
the standard quantum tomography. Instead of  discussing the errors
of measured projections separately, one can make a synthesis of all
the results by forming an  estimate of the unknown quantum state, and then
evaluate the average error. Let us apply this strategy  to the
detection of a complete set of  MCO.  A generic quantum state 
may be decomposed  in the basis of MCO as follows
\begin{equation} \label{compl_decomp}
\bar{\rho}=\sum_{\alpha j}\bar{w}_{\alpha j} \Pi_{\alpha j},
\end{equation}
where $\Pi_{\alpha j}$ are projectors on the eigenspaces of
a complete set of MCO, the coefficients $\bar{w}_{\alpha j}$ being
determined by the true probabilities
$p_{\alpha j}={\rm Tr}\{\bar\rho\Pi_{\alpha j}\}$ as follows 
\cite{king},
\begin{equation} \label{w}
\bar{w}_{\alpha j}=p_{\alpha j}-\frac{1}{p+1},
\end{equation}

Now suppose that the complementary observations are done
with $N$ particles each. Due to
fluctuations, the registered relative frequencies
$f_{\alpha j}$ will
generally differ from the true probabilities $p_{\alpha j}$
and so the estimated state
\begin{equation} \label{est_decomp}
\rho=\sum_{\alpha j}w_{\alpha j} \Pi_{\alpha j}.
\end{equation}
Let us see how much. Perhaps the most simple estimation strategy
is the direct inversion
based on the above mentioned independence of MCO,
\begin{equation} \label{estimation}
w_{\alpha j}=f_{\alpha j} -\frac{1}{p+1}.
\end{equation}
The error will be quantified by evaluating the Hilbert-Schmidt
distance between the true and the estimated state
\begin{equation} \label{dist}
d={\rm Tr}\{(\rho-\bar\rho)^2\}.
\end{equation}
The mean distance (error) is then given by averaging $d$ over many
repetitions of the estimation procedure, each yielding 
slightly different estimates of $\bar{w}_{\alpha j}$,
\begin{equation} \label{error_est}
E_{\rm est}=\langle d\rangle.
\end{equation}
Using Eqs.(\ref{compl_decomp})-(\ref{dist}) and
the definition of complementarity (\ref{def-compl})
in Eq.(\ref{error_est}) we get,
\begin{equation} \label{e1}
\langle d\rangle=\sum_{\alpha j}
\langle(\Delta w_{\alpha j})^2\rangle=
\frac{1}{N}\sum_{\alpha j}
p_{\alpha j}(1-p_{\alpha j}).
\end{equation}
Notice that for the given number of input particles the mean
distance of the estimated state from the true state is proportional
to the total error (\ref{error}), which depends on $\bar\rho$ only
through ${\rm Tr}\bar\rho^2$. This is one of the key results of
this contribution. It  shows an alternative interpretation of the
total error $E$ and the invariant information: The total lack of
the information as defined by Brukner and Zeilinger determines the
mean error of the standard reconstruction based on the measurement
of a complete set of MCO.

Given the same data, the accuracy of our estimate based on them 
strongly depends on the chosen reconstruction procedure. Direct 
inversion
(\ref{estimation}) is simple and straightforward because it
implicitly uses the apparent statistical independence of MCO. But
this  is not the only possibility there. Keeping in mind than we
want to characterize the information gained through the measurement
one should use the best reconstruction method available. 
To evaluate the error of the optimal estimation procedure it is 
more convenient to decompose the true density matrix in a basis of 
{\em orthogonal} observables rather than MCO,
\begin{equation}\label{true_dec_orto}
\bar\rho=\frac{1}{p}+\sum_k \bar a_k \Gamma_k,
\end{equation}
where the $p^2-1$ unknown parameters $\bar a_k$ provide a minimal
representation of the state $\bar\rho$, and $\Gamma_j$ are
orthonormal: ${\rm Tr}\{\Gamma_j\Gamma_k\}=\delta_{jk}$.
When the measurement is over an estimate $\rho$ of the true state
is formed,
\begin{equation}\label{est_dec_orto}
\rho=\frac{1}{p}+\sum_k a_k(f_k) \Gamma_k.
\end{equation}
The parameters $a_k$ specifying the estimated state
depend on the registered frequencies according to the
given reconstruction procedure.
Now we calculate the mean distance between the estimated and true
states. Trivial calculation shows that it
is given by a sum of the variances of the estimated parameters,
\begin{equation}\label{dist_orto}
\langle d \rangle=\sum_k \langle (\Delta a_k)^2\rangle.
\end{equation}
To proceed, the variances appearing in (\ref{dist_orto})
must somehow be determined. This may be a
nontrivial problem because generally the estimated parameters
$a_k$ might depend on the measured frequencies $f_k$ in a very
complicated way. However, for our purpose, it is enough to evaluate
the performance of the optimum estimation.
It was shown that the variances of estimated parameters
cannot be less than the Cramer-Rao lower bound \cite{rao}.
Further, it is known that maximum-likelihood estimators
attain the bound asymptotically (for large $N$) \cite{ml-efficient}.
In our case the Cramer-Rao bound reads:
\begin{equation}\label{CRLB}
\langle (\Delta a_k)^2\rangle\ge {\rm F}^{-1}_{kk},
\end{equation}
where ${\rm \bf F}$ is the Fisher information matrix defined as
\cite{helstrom,frieden}
\begin{equation}\label{fish-def}
{\rm F}_{kl}=\left\langle
\frac{\partial}{\partial
a_k}\log P({\bf n}|{\bf a})
\frac{\partial}{\partial a_l}\log P({\bf n}|{\bf a})
\right\rangle.
\end{equation}
Here $P({\bf n}|{\bf a})$ denotes the probability
of registering data ${\bf n}$
provided the true state is ${\bf a}$, 
and the averaging is done over
the registered data.
For large $N$, one is allowed to replace the multinomial distribution 
$P$ by its Gaussian approximation. Keeping the notation of MCO 
registrations we have, 
\begin{equation}\label{gauss}
\log P({\bf n}|{\bf a})\approx
-N^2\sum_{\alpha j}\frac{(f_{\alpha j}-p_{\alpha j})^2}
{\sigma^2_{\alpha j}},
\end{equation}
where $\sigma^2_{\alpha j}=N p_{\alpha j}(1-p_{\alpha j})$
as before. Under this approximation the Fisher matrix becomes,
\begin{equation}\label{F}
{\rm F}_{kl}= N^2\sum_{\alpha j}
\frac{{\rm Tr}\{\Gamma_k\Pi_{\alpha j}\}
{\rm Tr}\{\Gamma_l\Pi_{\alpha j}\}}
{\sigma^2_{\alpha j}}.
\end{equation}
The main formal result of the paper is obtained from Eqs.~(\ref{dist_orto}) 
and (\ref{CRLB}): Operational information defined  as the
mean error of the optimal estimation from the measurement of the
chosen set of observables is given by the trace 
of the inverse of the Fisher matrix,
\begin{equation}\label{dist-fish}
\langle d\rangle_{\rm opt}={\rm Sp}\,{\rm \bf F}^{-1}.
\end{equation} 
It quantifies the maximum amount of information
about an unknown state gained by the measurement performed.
Eq.~(\ref{dist-fish}) has the following simple geometrical 
interpretation: 
On registering counts ${\bf n}$, the probability (\ref{gauss}), 
considered now as a function of ${\bf a}$, 
characterizes the likelihood of various states ${\bf a}$. 
In terms of the Fisher matrix it reads:
\begin{equation}\label{quadratic}
\log{\cal L}({\bf a})\approx -\sum_{kl}(a_k-\tilde{a}_k)(a_l-\tilde{a}_l)
F_{kl},
\end{equation}
where $\tilde {\bf a}$ specifies the Maximum-likelihood solution.
Let us define the error volume as the set of density matrices
whose likelihoods do not drop below a certain threshold, 
${\cal L}\ge\rm{const}$. According to (\ref{quadratic})
the error volume is an ellipsoid, the lengths of its axes being
inversely proportional to the eigenvalues of the Fisher matrix.
The optimal estimation error $\langle d\rangle_{\rm opt}$ 
can thus be interpreted as the sum of half-axes of the 
error ellipsoid which corresponds to the chosen 
set of measurement. From this point of view, the synthesis of
all the quantum observations is equivalent to the registration
of the orthogonal observables $\Gamma'$ defining those axes.
In this new representation the Fisher matrix attains
the diagonal form, which means that the estimates of the transformed 
quantum-state ``coordinates'' $\bar{a}'_k={\rm Tr}\{\bar\rho \Gamma'\}$ 
fluctuate independently. Their statistical independence 
justifies the adding  their variances as it appears 
in Eq.~(\ref{dist-fish}).

This geometrical construction also shows that the structure
of uncertainties related to the quantum tomography is richer than
what a single number (\ref{dist-fish}) might indicate. 
The chosen measurement might provide 
different resolutions in different directions, depending on 
the shape and orientation of the noise ellipsoid.
Let us mention in passing that the resolution
optimized over all measurements has been shown to 
provide a natural statistical distance 
between two quantum states \cite{stat-dist}.

Now, we will go back to the measurement of a complete set of MCO. 
It is not difficult to see that the invariance of the error (\ref{e1})
of the direct inversion (\ref{estimation}) stems from the fact that
the projectors measured are considered as statistically independent
observables. Each parameter $a_k$ in the decomposition
(\ref{est_decomp}) is determined by measuring only one 
projector --- the rest contributes nothing. 
As will be shown below, such data treatment cannot be optimal. 
This leads us to the following important question: 
Will the invariance of the estimation error survive if the direct
inversion is replaced by the optimum data treatment? It turns out
that the answer depends on the dimension $p$ of the Hilbert space. In
the simplest non trivial case of $p=2$ the answer is in the
affirmative. However for $p\ge 3$ a deviation appears. Let us have
a look at the behavior of $\langle d \rangle_{\rm opt}$ for $p=3$.
Straightforward calculation of ${\rm Tr}\,{\bf F}^{-1}$ in
this case yields:
\begin{equation}\label{fishp3}
\langle d\rangle_{\rm opt}=\frac{1}{N^2}
\sum_{\alpha j} \sigma_{\alpha j}^2-\frac{1}{N^2}
\sum_{\alpha} \frac{\sigma_{\alpha 1}^4+
\sigma_{\alpha 2}^4+\sigma_{\alpha 3}^4}
{\sigma_{\alpha 1}^2+\sigma_{\alpha 2}^2+\sigma_{\alpha 3}^2}.
\end{equation}
Notice that the first expression on the right-hand side is
exactly the invariant total error (\ref{e1}).
The second term then quantifies the improvement of the optimum
estimation upon the simple linear inversion
(\ref{estimation}). One can easily check that this term is not 
an invariant quantity. So it turns out that the error of the optimum
estimation from the measurement of a complete set
of MCO is not invariant with respect to unitary transformations.
This means that the maximum amount of
information about the true state that can be gained from
such measurements might differ even for true states
of the same degree of purity! One pays for the optimality of
the reconstruction procedure by loosing its
invariant character (universality). Or conversely,
one can have an invariant reconstruction at the expense
of loosing precision.

Let us now address the broader aspects of optimal information
processing discussed above. It seems that the non-invariant
character of $\langle d\rangle_{\rm opt}$ originates from the
``inhomogeneity'' of the measured quorum of MCO. Two projectors
drawn from the quorum can be either ``complementary'', ${\rm
Tr}\{\Pi_1 \Pi_2\}=1/p$, or orthogonal, ${\rm Tr}\{\Pi_1
\Pi_2\}=0$, to each other.
Consequently, the mesh established by MCO is not as regular as it
might have seemed at the beginning. 
Provided that this  is ignored, the observations may be regarded
as ``independent'' on average, but such treatment 
is obviously not optimal.  Some additional information can be gained 
provided those observations are treated as noise dependent 
observations \cite{maxlik}.

The invariance of the optimum estimation error can be restored provided a
more ``homogeneous'' quorum of observables is measured. This can be
illustrated on the thought measurement of the following 
observables,
\begin{equation}\label{orto-quorum}
\Pi_j=\frac{1}{2}+\frac{\sqrt{2}}{2}\Gamma_j,\quad j=1,..,p^2-1,
\end{equation}
where the normalization and the factor of $1/2$ ensure that
$0<\Pi_j<1$, so that $\Pi_j$ are actually POVM elements
which can be associated probabilities
\begin{equation}\label{prob-orto}
p_j=\frac{1}{2}+\frac{\sqrt{2}}{2} \bar a_j,
\end{equation}
and $\bar a_j$ specify the true state as before.
Because we now have ${\rm Tr}\{\Pi_j\Gamma_k\}=\delta_{jk}$,
the Fisher matrix (\ref{F}) assumes the diagonal form.
A simple calculation then yields
\begin{equation}\label{inv-fish2}
\langle d\rangle_{\rm opt}=\frac{2}{N^2}\sum_j \sigma^2_j.
\end{equation}
Notice that this expression resembles the one for the 
invariant total error (\ref{error}).
The mean error of orthogonal
observations is obtained by substituting
$\sigma^2_j=N p_j(1-p_j)$ and Eq.~(\ref{prob-orto})
in Eq.~(\ref{inv-fish2}),
\begin{equation}\label{error-orto}
N\langle d\rangle_{\rm ortho}=\frac{p^2-1}{2}+
\frac{1}{p}-{\rm Tr}\bar\rho^2.
\end{equation}
This is, obviously, an invariant quantity.

Finally, let us mention that optimal quantum reconstruction
schemes provide invariant estimation errors.
Such an optimal measurement would consist in measuring
the ``unknown'' quantum state along its eigenvectors,
$\bar{\rho}\Pi_j=\lambda_j\Pi_j$.
It would obviously give mutually exclusive
results due to their orthogonality --- Maximum likelihood
estimation would always coincide with a deterministic data 
inversion. This gives the minimal achievable value of the 
average error: $N E_{opt}=   \sum_j \lambda_j(1-\lambda_j) =  
1-{\rm Tr}\bar\rho^2$. 
It should be clear that in the case of a generic quantum 
state whose diagonalizing basis 
is unknown and should be estimated
together with $\{\lambda_j\}$, the optimal information can only 
be attained asymptotically, in the limit of 
the large signal to noise ratio 
$N\rightarrow \infty$ \cite{adaptive}. 

To conclude, we have shown that
the invariant information introduced by
Brukner and Zeilinger is related to the problem of the
estimation of a quantum state.
It quantifies how the estimated state differs on average from
the true states in the sense of the Hilbert-Schmidt norm.
It depends on the quality of the measurement and on the 
data treatment adopted. Usually, the
measurement of a complete set of mutually complementary observables
is assumed. In this case the measured state may be easily
found via a linear inversion. The average distance
between the estimated and true state is given
directly by the invariant information.

The amount of extracted information may be increased provided that
data are used in optimal way achieving the ultimate limit given by
(the inverse of) the Fisher information. When doing this the unitary
invariance is sacrificed in favor of the accuracy of estimation.
Consequently, some ``directions'' can be identified
easier than another. This seems to be just an example of the
interplay between the universality and effectiveness of a particular
method.

Finally, we touched the the concept of mutually complementary observables.
One advantage of such observations is that the data can be more easily
processed to get information about signal via the direct inversion. This is
possible, unitary invariant but not efficient --- one can do something better.
The inefficiency of such an approach can be recognized in the structure of
the Brukner-Zeilinger information. In the full spectrum of mutually
complementary observables some projections commute but some do not. 
However, their variances are always added regardless of the commutativity or
non-commutativity! This is not optimal. In the light of this, 
the often discussed issue of the existence or
non-existence of a complete set of  mutually complementary observables in
some dimensions seems to be just an academic question without 
essential consequences for the problem of the state determination.

We would like to thank \v{C}. Brukner, A. Zeilinger and
B.-G. Englert for pertinent comments. This work was supported by
grants No. LN00A015 and J14/98 of the Czech Ministry of Education, 
by the TMR Network ERB-FMRX-CT96-0057 of the EU, and 
by the East-West program of the Austrian Government.


\begin{references}

\bibitem{zeilinger} A. Zeilinger, Found. Phys. {\bf 29},
631 (1999).

\bibitem{caslav}  C. Brukner, A. Zeilinger, 
Phys. Rev. Lett. {\bf 83}, 3354 (1999).

\bibitem{malus} \v{C}. Brukner, A. Zeilinger,
Acta Phys. Slov. {\bf 49}, 647 (1999).

\bibitem{essence}
\v{C}. Brukner, M. Zukowski and A. Zeilinger, eprint arXiv:quant-ph/0106119.

\bibitem{ivanovic} I.D. Ivanovi\'{c}, J. Phys. A: Math. Gen. {\bf 14},
3241 (1981).

\bibitem{schwinger} J.~Schwinger, ``Quantum Mechanics,''
Edited by B.-G.~Englert (Springer-Verlag, Berlin, 2001).

\bibitem{wf-mco}
W. K. Wootter and B. D. Fields, Ann. Phys. (N.Y.) {\bf 191}, 363
(1989); see also \cite{positivity}.

\bibitem{positivity} To be precise, a measurement of one such 
an observable usually gives {\em some} information 
about observables complementary to it because of the positivity constraint 
which the density operator must obey. However, all the possible values of 
those observables that are consistent with this constraint are equally
likely; in this sense, MCO are unbiased.

\bibitem{king}
B.-G. Englert and Y. Aharonov, Phys. Lett. A {\bf 284}, 1 (2001).

\bibitem{rao} C.R. Rao, Bull. Calcutta Math. Soc. {\bf 37},
81 (1945); H. Cram\'er, ``Mathematical methods of
statistics" (Princeton University Press, 1946).

\bibitem{ml-efficient} R.A. Fisher, Proc. Camb. Phi. Soc. {\bf 22},
700 (1925).

\bibitem{helstrom} C.W. Helstrom, 
``Quantum Detection and Estimation Theory,''
(Academic Press, New York, 1976).

\bibitem{frieden} B.R. Frieden, 
``Physics from Fisher Information: A Unification,''
(Cambridge University Press, 1999).

\bibitem{stat-dist} W.K. Wootters, Phys. Rev. {\bf  D 23}, 
357 (1981); Braunstein and Caves, 
Phys. Rev. Lett. {\bf 72}, 3439 (1994).

\bibitem{maxlik}  
Z. Hradil, Phys. Rev. A {\bf 55}, R1561 (1997);
J.~\v{R}eh\'{a}\v{c}ek, Z.~Hradil, and M.~Je\v{z}ek,
Phys. Rev. A {\bf 62}, 040303(R) (2000).

\bibitem{adaptive} Part of the ensemble ($\propto \sqrt{N}$) 
is spent on finding the diagonalizing basis,
the rest is used for estimating $\{\lambda_j\}$.





\end{references}
\end{document}